\documentclass[aps,prc,preprint,amsmath,showpacs,superscriptaddress,nofootinbib]{revtex4}
\usepackage{graphicx,epsfig,psfrag}
\newcommand{\beqy}{\begin{eqnarray}}
\newcommand{\eeqy}{\end{eqnarray}}
\newcommand{\bmlet}{\begin{subequations}}
\newcommand{\emlet}{\end{subequations}}

\begin{document}

\title{Properties of the outer crust of strongly magnetized neutron stars from Hartree-Fock-Bogoliubov atomic mass models}

\author{N. Chamel}
\affiliation{Institute of Astronomy and Astrophysics, Universit\'e Libre de Bruxelles, CP 226, Boulevard du Triomphe, B-1050 Brussels, Belgium}
\author{R. L. Pavlov}
\affiliation{Institute for Nuclear Research and Nuclear Energy, Bulgarian Academy of Sciences, 72 Tsarigradsko Chaussee, 1784 Sofia, Bulgaria}
\author{L. M. Mihailov}
\affiliation{Institute of Solid State Physics, Bulgarian Academy of Sciences, 72 Tsarigradsko Chaussee, 1784 Sofia, Bulgaria}
\author{Ch. J. Velchev}
\affiliation{Institute for Nuclear Research and Nuclear Energy, Bulgarian Academy of Sciences, 72 Tsarigradsko Chaussee, 1784 Sofia, Bulgaria}
\author{Zh. K. Stoyanov}
\affiliation{Institute for Nuclear Research and Nuclear Energy, Bulgarian Academy of Sciences, 72 Tsarigradsko Chaussee, 1784 Sofia, Bulgaria}
\author{Y. D. Mutafchieva}
\affiliation{Institute for Nuclear Research and Nuclear Energy, Bulgarian Academy of Sciences, 72 Tsarigradsko Chaussee, 1784 Sofia, Bulgaria}
\author{M. D. Ivanovich}
\affiliation{Institute for Nuclear Research and Nuclear Energy, Bulgarian Academy of Sciences, 72 Tsarigradsko Chaussee, 1784 Sofia, Bulgaria}
\author{J.M. Pearson}
\affiliation{D\'ept. de Physique, Universit\'e de Montr\'eal, Montr\'eal (Qu\'ebec), H3C 3J7 Canada}
\author{S. Goriely}
\affiliation{Institute of Astronomy and Astrophysics, Universit\'e Libre de Bruxelles, CP 226, Boulevard du Triomphe, B-1050 Brussels, Belgium}

\begin{abstract}
The equilibrium properties of the outer crust of cold nonaccreting magnetars (i.e. neutron stars endowed with very strong magnetic 
fields) are studied using the latest experimental atomic mass data complemented with a microscopic atomic mass model 
based on the Hartree-Fock-Bogoliubov method. The Landau quantization of electron motion caused by the strong magnetic field
is found to have a significant impact on the composition and the equation of state of crustal matter. It is also shown that 
the outer crust of magnetars could be much more massive than that of ordinary neutron stars. 
\end{abstract}


\maketitle

\section{Introduction}
\label{intro}

Neutron stars are among the most strongly magnetized objects in the universe~\cite{hae07}. Radio 
pulsars are endowed with typical surface magnetic fields of order $10^{12}$~G~\cite{seir04}. 
A few radio pulsars have been found to have significantly higher surface magnetic fields of 
order $10^{13}-10^{14}$~G~\cite{ka11}. Surface magnetic fields up to $2.4\times 10^{15}$~G have 
been inferred in soft-gamma ray repeaters (SGRs) and anomalous x-ray pulsars (AXPs) from both spin-down 
and spectroscopic studies~\cite{mer08,mcgill}. Even stronger fields might exist in the interior of 
these neutron stars, as suggested by various observations~\cite{ste05,kam07,rea10}. Duncan and Thompson 
showed that strong magnetic fields up to $\sim 10^{16}-10^{17}$~G can be generated via dynamo effects 
in hot newly-born neutron stars with initial periods of a few milliseconds~\cite{td93} leading to the 
formation of strongly magnetized neutron stars thus dubbed \emph{magnetars} (see e.g. Ref.~\cite{wt06} for 
a review). 
Numerical simulations confirmed that magnetic fields of order $\sim 10^{15}-10^{16}$~G ccan be produced 
during supernovae explosions due to the magnetorotational instability~\cite{ard05}. 
A very amount of magnetic energy can be occasionally released in crustquakes thus triggering the 
gamma-ray bursts observed in SGRs and AXPs~\cite{td95}. This scenario has been recently supported by the 
detection of quasiperiodic oscillations (QPOs) in the x-ray flux of giant flares from a few SGRs. Some of 
these QPOs coincide reasonably well with seismic crustal modes thought to arise from the release of magnetic 
stresses~\cite{ws07,sa07}. The huge luminosity variation suggests $B\gtrsim 10^{15}$~G at the star surface 
thus lending support to the magnetar hypothesis~\cite{vie07}. According to the virial theorem, the upper 
limit on the neutron-star magnetic fields is of the order of $10^{18}$~G~\cite{lai91}.
This limit has been confirmed by numerical magnetohydrodynamics simulations~\cite{boc95,car01,kiu08}.

In this paper, we study the impact of a strong magnetic field on the equilibrium properties of the outer crust 
of cold non-accreting neutron stars along the lines of Ref.~\cite{lai91}. For this purpose we made use of the 
most recent experimental atomic mass data complemented with a theoretical atomic mass table based on the Hartree-Fock-Bogoliubov 
(HFB) method~\cite{gcp10}. In Sec.~\ref{model}, we present the microscopic model used to describe the outer crust of a 
magnetar. Results are discussed in Sec.~\ref{results} and simple analytical formulas are derived 
in the limit of strongly quantizing fields in Sec.~\ref{strong}.

\section{Microscopic model of magnetar crusts}
\label{model}

In the magnetar theory, neutron stars are born with very strong magnetic fields of order $B\sim 10^{16}-10^{17}$~G 
which decay on a typical time scale of order $\gtrsim 10^3$ years~\cite{dss09}. We assume that the magnetic fields 
are sustained long enough to alter the formation of neutron-star crusts. In the model we adopt here~\cite{lai91}, the 
neutron-star crust is assumed to be made of ``cold catalyzed matter'', i.e.  matter in its ground state at zero temperature 
and in a uniform magnetic field. The magnetic field mostly affects the outermost region of the crust where atoms are supposed 
to be fully ionized and arranged in a body centered cubic lattice~\cite{lrr}. We determined the equilibrium composition of 
each layer of the outer crust at a given pressure $P$ by minimizing the Gibbs free energy per nucleon 
\begin{equation}
\label{1}
g=\frac{\mathcal{E}+P}{n}
\end{equation}
where $\mathcal{E}$ is the average energy density and $n$ the average nucleon number density. 
Assuming that each layer of the outer crust contains only one nuclear species with proton number 
$Z$ and atomic number $A$, the average energy density can be expressed as
\begin{equation}
\label{2}
\mathcal{E}=n_N M^\prime(Z,A)+\mathcal{E}_e+\mathcal{E}_L
\end{equation}
where $n_N=n/A$ is the number density of nuclei, $M^\prime(Z,A)$ their mass (including the rest mass of nucleons 
and $Z$ electrons), $\mathcal{E}_e$ the energy density of electrons after subtracting out the electron rest mass energy density 
and $\mathcal{E}_L$ the lattice energy density. The nuclear mass $M^\prime(Z,A)$ can be obtained from the atomic mass $M(Z,A)$ 
after substracting out the binding energy of the atomic electrons
(see Eq.~(A4) of Ref.~\cite{lpt03})
\begin{equation}
\label{3}
M^\prime(A,Z) = M(A,Z) + 1.44381\times 10^{-5}\,Z^{2.39} + 
1.55468\times 10^{-12}\,Z^{5.35}
\end{equation}
where both masses are expressed in units of MeV. As in Ref.~\cite{lai91}, we will ignore the effects of the magnetic field on 
nuclear masses. 
Shell correction calculations using the simple Nilsson model predict that magnetic fields  $\sim 10^{16}$~G can change 
nuclear shell structure hence also nuclear masses~\cite{kon00,kon01}. However, a very recent study based on fully self-consistent 
relativistic mean-field calculations concluded that significantly higher fields $B\gtrsim 10^{17}$~G are required to 
affect substantially the composition of the outer crust~\cite{pen11}. 
In Ref.~\cite{lai91}, the authors used the experimental atomic 
masses from Ref.~\cite{wap76} supplemented with the mass model of Ref.~\cite{wap77}. In this paper, we made use of the 
most recent experimental atomic mass data from a preliminary unpublished version of an updated Atomic Mass Evaluation 
(AME)~\cite{audi11}. For the masses that have not yet been measured, we employed the microscopic atomic mass model HFB-21 of 
Ref.~\cite{gcp10} based on the HFB method using a generalized Skyrme effective nucleon-nucleon interaction~\cite{cgp09} 
supplemented with a microscopic contact pairing interaction~\cite{cha10}. The parameters of the Skyrme interaction BSk21 underlying the 
HFB-21 model were fitted to the 2149 measured masses of nuclei with $N$ and $Z \ge 8$ given in the 2003 AME~\cite{audi03}. For this it 
was necessary to add two phenomenological corrections to the HFB ground-state energy: (i) a Wigner energy (which contributes significantly 
only for light nuclei or nuclei with $N$ close to $Z$) and (ii) a correction for the spurious rotational and vibrational collective energies. 
With an rms deviation as low as 0.58 MeV, this atomic mass model is well-suited for describing the neutron-rich nuclei found in the 
outer crust of a neutron star. Incidentally, the parameters of the Skyrme interaction were simultaneously constrained to 
reproduce the zero-temperature equation of state of homogeneous neutron matter, as determined by many-body calculations with realistic 
two- and three-nucleon forces~\cite{ls08}, from very low densities up to the maximum density found in stable neutron stars. 
For this reason, the Skyrme interaction BSk21 could be reliably extrapolated beyond the outer crust thus providing a unified description 
of all regions of a neutron star. In particular, this interaction has been recently used to determine the equation of state of cold 
non-accreting non-magnetized neutron stars~\cite{pea11,pcgc12} and has been found to be compatible with measurements of neutron-star 
masses~\cite{cfpg11}.

In the presence of a strong magnetic field, the electron motion perpendicular to the field is quantized into Landau levels
(see for instance Ref.~\cite{hae07}). 
For sufficiently strong fields, the electron cyclotron energy becomes comparable to the electron rest-mass energy. This happens 
for $B>B_c$ where the critical magnetic field $B_c$ is given by 
\begin{equation}
\label{4}
B_c=\frac{m_e^2 c^3}{e\hbar}\simeq 4.4\times 10^{13}\, \rm G\, .
\end{equation}
Surface magnetic fields $B>B_c$ have been inferred in various kinds of neutron stars~\cite{mer08,ka11}. 
Ignoring electron polarization effects (see e.g. Chap. 4 in Ref.~\cite{hae07} and references therein) and 
treating electrons as a relativistic Fermi gas, the energies of 
Landau levels (which were actually first found by Rabi as early as 1928~\cite{rab28}) are given by
\begin{equation}
\label{5}
e_{\nu} = \sqrt{c^2 p_z^2+m_e^2 c^4(1+2\nu B_\star)}
\end{equation}
\begin{equation}
\label{6}
\nu = n_L + \frac{1}{2}+\sigma\, ,
\end{equation}
where $n_L$ is any non-negative integer, $\sigma=\pm 1/2$ is the spin, $p_z$ is the component of the momentum along the field, and 
$B_\star=B/B_c$. The electron anomalous magnetic moment is small and has been neglected. 
For a given magnetic field strength $B_\star$, The number of occupied Landau levels is determined by the electron number density $n_e$ 
\begin{equation}
\label{7}
n_e =\frac{2 B_\star}{(2 \pi)^2 \lambda_e^3} \sum_{\nu=0}^{\nu_{\rm  max}} g_\nu x_e(\nu)\, ,
\end{equation}
\begin{equation}
\label{8}
x_e(\nu) =\sqrt{\gamma_e^2 -1-2 \nu B_\star}\, ,
\end{equation}
where $\lambda_e=\hbar/m_e c$ is the electron Compton wavelength, $\gamma_e$ is the electron chemical potential in units of 
the electron rest mass energy, that is,
\begin{equation}
\label{9}
\gamma_e =\frac{\mu_e}{m_e c^2}\, ,
\end{equation}
while the degeneracy $g_\nu$ is $g_\nu=1$ for $\nu=0$ and $g_\nu=2$ for $\nu\geq 1$. 

The electron energy density $\mathcal{E}_e$ and corresponding electron pressure $P_e$ are given by (see e.g.  Ref.~\cite{lai91} 
and references therein)
\begin{equation}
\label{10}
\mathcal{E}_e=\frac{B_\star m_e c^2}{(2 \pi)^2 \lambda_e^3} \sum_{\nu=0}^{\nu_{\rm max}}g_\nu(1+2\nu B_\star) \psi_+ \biggl[\frac{x_e(\nu)}{\sqrt{1+2\nu B_\star}}\biggr]-n_e m_e c^2\, ,
\end{equation}
and 
\begin{equation}
\label{11}
P_e=\frac{B_\star m_e c^2}{(2 \pi)^2 \lambda_e^3} \sum_{\nu=0}^{\nu_{\rm max}}g_\nu(1+2\nu B_\star) \psi_- \biggl[\frac{x_e(\nu)}{\sqrt{1+2\nu B_\star}}\biggr]\, ,
\end{equation}
respectively, where
\begin{equation}
\label{12}
\psi_\pm(x)=x\sqrt{1+x^2}\pm\ln(x+\sqrt{1+x^2})\, .
\end{equation}

In the absence of magnetic fields $B=0$, the electron energy density and pressure reduce to (see e.g. Chap. 2 in Ref.~\cite{hae07})
\begin{equation}
\label{10b}
\mathcal{E}_e=\frac{m_e c^2}{8 \pi^2 \lambda_e^3} \biggl[x_r(1+2x_r^2)\sqrt{1+x_r^2}-\ln(x_r+\sqrt{1+x_r^2})\biggr]-n_e m_e c^2\, ,
\end{equation}
and 
\begin{equation}
\label{11b}
P_e=\frac{m_e c^2}{8 \pi^2 \lambda_e^3} \biggl[x_r\left(\frac{2}{3}x_r^2-1\right)\sqrt{1+x_r^2}+\ln(x_r+\sqrt{1+x_r^2})\biggr]\, ,
\end{equation}
respectively, where $x_r=\hbar (3\pi^2 n_e)^{1/3}/(m_e c)$ is the relativity parameter.

According to the Bohr-van Leeuwen theorem~\cite{bvl32}, the lattice energy density is not affected by the magnetic field 
(we neglect here the small contribution due to the quantum zero-point motion of ions~\cite{bai09}). 
For point-like ions arranged in a body centered cubic lattice, the lattice energy density is approximately given by~\cite{col60}
\begin{equation}
\label{13}
\mathcal{E}_L=-1.44423 Z^{2/3} e^2 n_e^{4/3}\, ,
\end{equation}
and the associated pressure is 
\begin{equation}
\label{14}
P_L=\frac{1}{3}\mathcal{E}_L\, .
\end{equation}
The total pressure $P$ is therefore
\begin{equation}
\label{15}
P=P_e+P_L\, .
\end{equation}

Equations (\ref{10}) through (\ref{15}) remain approximately valid at finite temperatures $T$ provided the Coulomb coupling parameter
$\Gamma\gg 10^2$ and the temperature $T$ is much smaller than the electron Fermi temperature $T_{\rm F}$. 

\section{Equilibrium composition and equation of state of magnetar crusts}
\label{results}

The equilibrium composition of the outer crust of a non-accreting magnetized neutron star at $T=0$ in a layer characterized by 
a pressure $P$ is determined by minimizing the Gibbs free energy per nucleon
\begin{equation}
\label{16}
g=\frac{\mathcal{E}+P}{n}=\frac{M^\prime(A,Z)}{A}+\frac{Z}{A}\left( \mu_e - m_e c^2 +\frac{4}{3}\frac{\mathcal{E}_L}{n_e}\right)\, .
\end{equation} 
Note that the value of $g$ at equilibrium is simply equal to the neutron chemical potential. 
Starting from the shallowest part of the crust where $P\sim 0$, we repeated the calculations by increasing the pressure until 
$g$ equals $m_n c^2$ ($m_n$ being the neutron mass) for some pressure $P_{\rm drip}$. The present model is not suited for describing 
the inner regions of the crust because neutrons drip out of nuclei for $P>P_{\rm drip}$ 
(for a study of the denser regions of strongly magnetized neutron stars, see e.g. Ref.~\cite{na11} for the inner crust and 
Refs.~\cite{bro11,sm11} for the core).

The results are summarized in Tables~\ref{tab1}--\ref{tab4}. 
For comparison, we also determined the composition of the outer crust in the absence of magnetic field.
Note that our results shown in Table~\ref{tab5} are slightly different from those given in Table III of Ref.~\cite{pea11} using the same 
HFB-21 atomic mass model because of our neglect of electron exchange and other small corrections that were included in Ref.~\cite{pea11}. 
Our results for $B_\star=0$ also differ from those obtained previously by the authors of Ref.~\cite{lai91} because of the use of more recent experimental 
and theoretical atomic mass data. In particular, the elements $^{124}$Ru and $^{118}$Kr that were found by the authors of Ref.~\cite{lai91} are now absent, 
whereas $^{79}$Cu, $^{80}$Ni, $^{124}$Sr and $^{121}$Y are present. Using the latest experimental mass tables the outer crust is found to 
contain nine nuclides with experimentally measured masses versus only six in the calculations of the authors of Ref.~\cite{lai91}. 

\begin{figure}[b]
\centering
\includegraphics[scale=0.5]{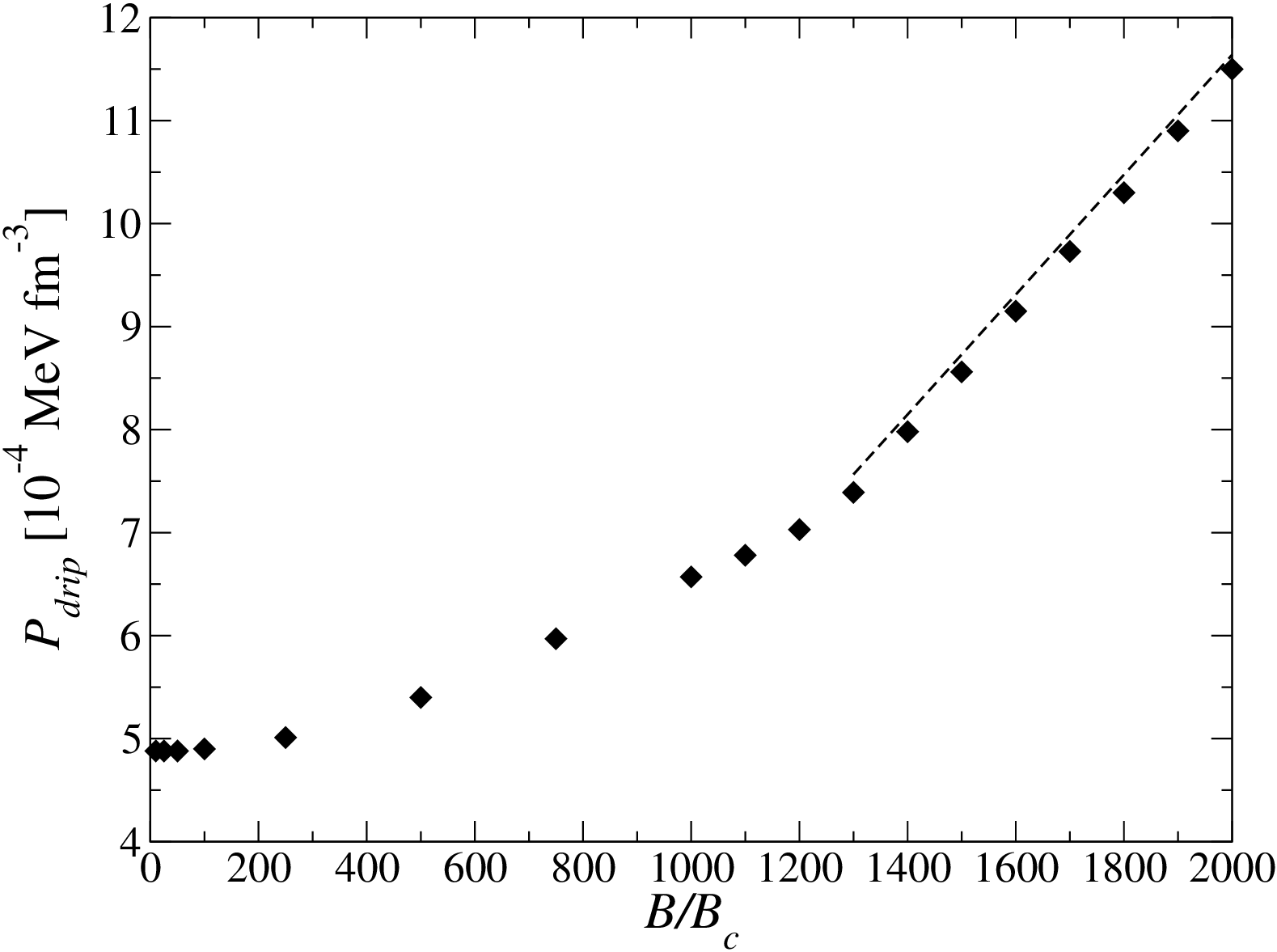}
\caption[]{Pressure $P_{\rm drip}$ at the neutron-drip transition in the crust of cold non-accreting neutron stars as a function 
of the magnetic field strength $B_\star=B/B_c$. The dashed line corresponds to the approximate expression~(\ref{41}) obtained in the 
strongly quantizing regime for which only the lowest level $\nu=0$ is filled.}
\label{fig1}
\end{figure}

For the ``weak'' magnetic fields prevailing in most pulsars $B_\star \lesssim 1$, the sequence of equilibrium nuclides in their outer 
crust is the same as that obtained in the absence of magnetic fields. However the highest density at which each nuclide can be found 
is increased, especially in the shallow region of the crust where the effects of Landau quantization are the most important. For 
instance, the maximum density at which $^{56}$Fe is found, is raised from $4.93\times 10^{-9}$ fm$^{-3}$ for $B_\star=0$ to 
$5.60\times 10^{-9}$ fm$^{-3}$ for $B_\star=1$. For the strong fields expected to exist in magnetars $B_\star\gg 1$, the sequence of 
equilibrium nuclides is changed. Table~\ref{tab6} indicates the magnetic field strength above which a nuclide appears or disappears. 
Moreover, strong magnetic fields tend to prevent neutrons from dripping out of nuclei. The pressure at the neutron drip transition thus 
increases from $4.88\times 10^{-4}$ MeV~fm$^{-3}$ for $B_\star=0$ to $1.15\times 10^{-3}$ MeV~fm$^{-3}$ for $B_\star=2000$, as shown in 
Fig.~\ref{fig1}. Note however that the equilibrium nuclide at the neutron-drip point remains $^{124}$Sr in all cases. 

\begin{table}
\centering
\caption{Composition and equation of state of the outer crust of strongly magnetized neutron stars
with $B_\star=B/B_c=1$. $n_{\rm min}$ ($n_{\rm max}$) is the minimum (respectively maximum) 
average baryon density (in units of fm$^{-3}$) at which the given nucleus is present. 
The pressure $P_{\rm max}$ (in units of MeV fm$^{-3}$) is the maximum pressure at which 
the given nucleus can be found. The surface density is estimated from Eq.~(\ref{25}).}
\label{tab1}
\vspace{.5cm}
\begin{tabular}{|cccccc|}
\hline
$Z$ & $N$ & $A$ & $n_{\rm min}$ &$n_{\rm max}$ & $P_{\rm max}$ \\
\hline
26& 30& 56& 2.50$\times 10^{-10}$ & 5.60$\times 10^{-9}$ & 4.15$\times 10^{-10}$ \\
28& 34& 62& 5.77$\times 10^{-9}$ & 1.60$\times 10^{-7}$& 4.22$\times 10^{-8}$\\
26& 32& 58& 1.61$\times 10^{-7}$ & 1.65$\times 10^{-7}$& 4.40$\times 10^{-8}$\\
28& 36& 64& 1.70$\times 10^{-7}$ & 8.01$\times 10^{-7}$& 3.56$\times 10^{-7}$\\
28& 38& 66& 8.27$\times 10^{-7}$ & 9.24$\times 10^{-7}$& 4.14$\times 10^{-7}$\\
36& 50& 86& 9.42$\times 10^{-7}$ & 1.86$\times 10^{-6}$& 1.03$\times 10^{-6}$\\
34& 50& 84& 1.92$\times 10^{-6}$ & 6.79$\times 10^{-6}$& 5.58$\times 10^{-6}$\\
32& 50& 82& 7.03$\times 10^{-6}$ & 1.67$\times 10^{-5}$& 1.77$\times 10^{-5}$\\
30& 50& 80& 1.74$\times 10^{-5}$ & 3.19$\times 10^{-5}$& 3.98$\times 10^{-5}$\\
29& 50& 79& 3.26$\times 10^{-5}$ & 4.35$\times 10^{-5}$& 5.87$\times 10^{-5}$\\
28& 50& 78& 4.45$\times 10^{-5}$ & 5.41$\times 10^{-5}$& 7.64$\times 10^{-5}$\\
28& 52& 80& 5.56$\times 10^{-5}$ & 8.09$\times 10^{-5}$& 1.24$\times 10^{-4}$\\
42& 82& 124& 8.37$\times 10^{-5}$ & 1.22$\times 10^{-4}$& 2.07$\times 10^{-4}$\\
40& 82& 122& 1.27$\times 10^{-4}$ & 1.48$\times 10^{-4}$& 2.55$\times 10^{-4}$\\
39& 82& 121& 1.51$\times 10^{-4}$ & 1.74$\times 10^{-4}$& 3.11$\times 10^{-4}$\\
38& 82& 120& 1.78$\times 10^{-4}$ & 1.95$\times 10^{-4}$& 3.53$\times 10^{-4}$\\
38& 84& 122& 1.99$\times 10^{-4}$ & 2.39$\times 10^{-4}$& 4.54$\times 10^{-4}$\\
38& 86& 124& 2.44$\times 10^{-4}$ & 2.56$\times 10^{-4}$& 4.87$\times 10^{-4}$\\
\hline
\end{tabular}
\end{table}

\begin{table}
\centering
\caption{Same as Table~\ref{tab1} for $B_\star=10$.}
\label{tab2}
\vspace{.5cm}
\begin{tabular}{|cccccc|}
\hline
$Z$ & $N$ & $A$ & $n_{\rm min}$ &$n_{\rm max}$ & $P_{\rm max}$ \\
\hline
26&	30&	56&	3.96$\times 10^{-9}$&	2.66$\times 10^{-8}$&	2.700$\times 10^{-9}$\\
28&	34&	62&	3.06$\times 10^{-8}$&	1.85$\times 10^{-7}$&	5.67$\times 10^{-8}$\\
28&	36&	64&	1.93$\times 10^{-7}$&	8.14$\times 10^{-7}$&	3.75$\times 10^{-7}$\\
28&	38&	66&	8.40$\times 10^{-7}$&	9.14$\times 10^{-7}$&	4.36$\times 10^{-7}$\\
36&	50&	86&	9.31$\times 10^{-7}$&	1.85$\times 10^{-6}$&	1.05$\times 10^{-6}$\\
34&	50&	84&	1.91$\times 10^{-6}$&	6.74$\times 10^{-6}$&	5.59$\times 10^{-6}$\\
32&	50&	82&	6.99$\times 10^{-6}$&	1.67$\times 10^{-5}$&	1.77$\times 10^{-5}$\\
30&	50&	80&	1.73$\times 10^{-5}$&	3.19$\times 10^{-5}$&	3.98$\times 10^{-5}$\\
29&	50&	79&	3.26$\times 10^{-5}$&	4.36$\times 10^{-5}$&	5.88$\times 10^{-5}$\\
28&	50&	78&	4.46$\times 10^{-5}$&	5.42$\times 10^{-5}$&	7.64$\times 10^{-5}$\\
28&	52&	80&	5.56$\times 10^{-5}$&	8.00$\times 10^{-5}$&	1.24$\times 10^{-4}$\\
42&	82&	124&	8.36$\times 10^{-5}$&	1.23$\times 10^{-4}$&	2.07$\times 10^{-4}$\\
40&	82&	122&	1.26$\times 10^{-4}$&	1.48$\times 10^{-4}$&	2.56$\times 10^{-4}$\\
39&	82&	121&	1.50$\times 10^{-4}$&	1.74$\times 10^{-4}$&	3.12$\times 10^{-4}$\\
38&	82&	120&	1.77$\times 10^{-4}$&	1.95$\times 10^{-4}$&	3.53$\times 10^{-4}$\\
38&	84&	122&	1.98$\times 10^{-4}$&	2.40$\times 10^{-4}$&	4.55$\times 10^{-4}$\\
38&	86&	124&	2.44$\times 10^{-4}$&	2.57$\times 10^{-4}$&	4.88$\times 10^{-4}$\\
\hline
\end{tabular}
\end{table}

\begin{table}
\centering
\caption{Same as Table~\ref{tab1} for $B_\star=100$.}
\label{tab3}
\vspace{.5cm}
\begin{tabular}{|cccccc|}
\hline
$Z$ & $N$ & $A$ & $n_{\rm min}$ &$n_{\rm max}$ & $P_{\rm max}$ \\
\hline
26&	30&	56&	6.28$\times 10^{-8}$&	2.84$\times 10^{-7}$&	2.97$\times 10^{-8}$ \\
28&	34&	62&	2.96$\times 10^{-7}$&	1.01$\times 10^{-6}$&	5.41$\times 10^{-7}$\\		
28&	36&	64&	1.04$\times 10^{-6}$&	1.68$\times 10^{-6}$&	1.47$\times 10^{-6}$\\
36&	50&	86&	1.76$\times 10^{-6}$&	2.33$\times 10^{-6}$&	2.62$\times 10^{-6}$\\
34&	50&	84&	2.40$\times 10^{-6}$&	7.58$\times 10^{-6}$&	7.34$\times 10^{-6}$\\
32&	50&	82&	7.85$\times 10^{-6}$&	1.73$\times 10^{-5}$&	1.97$\times 10^{-5}$\\
30&	50&	80&	1.80$\times 10^{-5}$&	3.17$\times 10^{-5}$&	4.19$\times 10^{-5}$\\
29&	50&	79&	3.24$\times 10^{-5}$&	4.41$\times 10^{-5}$&	6.10$\times 10^{-5}$\\
28&	50&	78&	4.51$\times 10^{-5}$&	5.48$\times 10^{-5}$&	7.85$\times 10^{-5}$\\
28&	52&	80&	5.62$\times 10^{-5}$&	8.04$\times 10^{-5}$&	1.27$\times 10^{-4}$\\
42&	82&	124&	8.38$\times 10^{-5}$&	1.23$\times 10^{-4}$&	2.10$\times 10^{-4}$\\
40&	82&	122&	1.27$\times 10^{-4}$&	1.49$\times 10^{-4}$&	2.58$\times 10^{-4}$\\
39&	82&	121&	1.51$\times 10^{-4}$&	1.75$\times 10^{-4}$&	3.14$\times 10^{-4}$\\
38&	82&	120&	1.78$\times 10^{-4}$&	1.95$\times 10^{-4}$&	3.55$\times 10^{-4}$\\
38&	84&	122&	1.99$\times 10^{-4}$&	2.40$\times 10^{-4}$&	4.57$\times 10^{-4}$\\
38&	86&	124&	2.44$\times 10^{-4}$&	2.57$\times 10^{-4}$&	4.90$\times 10^{-4}$\\
\hline
\end{tabular}
\end{table}

\begin{table}
\centering
\caption{Same as Table~\ref{tab1} for $B_\star=1000$.}
\label{tab4}
\vspace{.5cm}
\begin{tabular}{|cccccc|}
\hline
$Z$ & $N$ & $A$ & $n_{\rm min}$ &$n_{\rm max}$ & $P_{\rm max}$ \\
\hline
26&	30&	56&	9.96$\times 10^{-7}$&	2.62$\times 10^{-6}$&	1.98$\times 10^{-7}$\\
28&	34&	62&	2.71$\times 10^{-6}$&	1.10$\times 10^{-5}$&	6.23$\times 10^{-6}$\\
28&	36&	64&	1.14$\times 10^{-5}$&	1.42$\times 10^{-5}$&	1.01$\times 10^{-5}$\\
38&	50&	88&	1.45$\times 10^{-5}$&	1.55$\times 10^{-5}$&	1.16$\times 10^{-5}$\\
36&	50&	86&	1.60$\times 10^{-5}$&	2.60$\times 10^{-5}$&	3.21$\times 10^{-5}$\\
34&	50&	84&	2.69$\times 10^{-5}$&	3.88$\times 10^{-5}$&	6.81$\times 10^{-5}$\\
32&	50&	82&	4.03$\times 10^{-5}$&	5.22$\times 10^{-5}$&	1.16$\times 10^{-4}$\\
30&	50&	80&	5.43$\times 10^{-5}$&	6.54$\times 10^{-5}$&	1.69$\times 10^{-4}$\\	
29&	50&	79&	6.68$\times 10^{-5}$&	7.32$\times 10^{-5}$&	2.03$\times 10^{-4}$\\
28&	50&	78&	7.48$\times 10^{-5}$&	7.92$\times 10^{-5}$&	2.28$\times 10^{-4}$\\
28&	52&	80&	8.12$\times 10^{-5}$&	9.03$\times 10^{-5}$&	2.83$\times 10^{-4}$\\
42&	82&	124&	9.37$\times 10^{-5}$&	1.07$\times 10^{-4}$&	3.70$\times 10^{-4}$\\
40&	82&	122&	1.10$\times 10^{-4}$&	1.16$\times 10^{-4}$&	4.09$\times 10^{-4}$\\
39&	82&	121&	1.18$\times 10^{-4}$&	1.79$\times 10^{-4}$&	4.59$\times 10^{-4}$\\
38&	82&	120&	1.82$\times 10^{-4}$&	2.17$\times 10^{-4}$&	5.04$\times 10^{-4}$\\
38&	84&	122&	2.21$\times 10^{-4}$&	2.74$\times 10^{-4}$&	6.20$\times 10^{-4}$\\
38&	86&	124&	2.78$\times 10^{-4}$&	2.92$\times 10^{-4}$&	6.57$\times 10^{-4}$\\
\hline
\end{tabular}
\end{table}

\begin{table}
\centering
\caption{Same as Table~\ref{tab1} for $B_\star=0$.}
\label{tab5}
\vspace{.5cm}
\begin{tabular}{|cccccc|}
\hline
$Z$ & $N$ & $A$ & $n_{\rm min}$ &$ n_{\rm max}$ & $P_{\rm max}$ \\
\hline
26 & 30 & 56 & 0 & 4.93$\times 10^{-9}$ & 3.36$\times 10^{-10}$      \\
28 & 34 & 62 & 5.09$\times 10^{-9}$ & 1.59$\times 10^{-7}$ & 4.20$\times 10^{-8}$ \\
26 & 32 & 58 & 1.60$\times 10^{-7}$ & 1.65$\times 10^{-7}$ & 4.39$\times 10^{-8}$ \\
28 & 36 & 64 & 1.70$\times 10^{-7}$ & 7.99$\times 10^{-7}$ & 3.55$\times 10^{-7}$ \\
28 & 38 & 66 & 8.26$\times 10^{-7}$ & 9.22$\times 10^{-7}$ & 4.13$\times 10^{-7}$ \\
36 & 50 & 86 & 9.42$\times 10^{-7}$ & 1.86$\times 10^{-6}$ & 1.03$\times 10^{-6}$ \\
34 & 50 & 84 & 1.92$\times 10^{-6}$ & 6.79$\times 10^{-6}$ & 5.57$\times 10^{-6}$ \\
32 & 50 & 82 & 7.05$\times 10^{-6}$ & 1.67$\times 10^{-5}$ & 1.77$\times 10^{-5}$ \\
30 & 50 & 80 & 1.74$\times 10^{-5}$ & 3.18$\times 10^{-5}$ & 3.98$\times 10^{-5}$ \\
29 & 50 & 79 & 3.26$\times 10^{-5}$ & 4.35$\times 10^{-5}$ & 5.87$\times 10^{-5}$ \\
28 & 50 & 78 & 4.46$\times 10^{-5}$ & 5.42$\times 10^{-5}$ & 7.64$\times 10^{-5}$ \\
28 & 52 & 80 & 5.57$\times 10^{-5}$ & 7.99$\times 10^{-5}$ & 1.24$\times 10^{-4}$ \\
42 & 82 & 124 & 8.36$\times 10^{-5}$ & 1.23$\times 10^{-4}$ & 2.07$\times 10^{-4}$\\
40 & 82 & 122 & 1.27$\times 10^{-4}$ & 1.48$\times 10^{-4}$ & 2.55$\times 10^{-4}$\\
39 & 82 & 121 & 1.51$\times 10^{-4}$ & 1.74$\times 10^{-4}$ & 3.11$\times 10^{-4}$\\
38 & 82 & 120 & 1.78$\times 10^{-4}$ & 1.95$\times 10^{-4}$ & 3.53$\times 10^{-4}$\\
38 & 84 & 122 & 1.99$\times 10^{-4}$ & 2.39$\times 10^{-4}$ & 4.54$\times 10^{-4}$\\
38 & 86 & 124 & 2.44$\times 10^{-4}$ & 2.56$\times 10^{-4}$ & 4.86$\times 10^{-4}$ \\
\hline
\end{tabular}
\end{table}

\begin{table}
\centering
\caption{Magnetic field strength $B_\star=B/B_c$ for the appearance (+) or the disappearance (-) of a nuclide in the 
the outer crust of a cold non-accreting neutron star.}
\label{tab6}
\vspace{.5cm}
\begin{tabular}{|cc|}
\hline
Nuclide & $B_\star$ \\
$^{58}$Fe(-)& 9 \\
$^{66}$Ni(-)& 67 \\ 
$^{88}$Sr(+)& 859 \\
$^{126}$Ru(+)& 1118 \\
$^{128}$Pd(+)& 1120 \\
$^{78}$Ni(-)& 1120 \\
$^{80}$Ni(-)&1250\\
$^{64}$Ni(-)&1668\\
$^{79}$Cu(-)&1791\\
$^{130}$Cd(+)&1804\\
$^{132}$Sn(+)&1987\\
\hline
\end{tabular}
\end{table}

\begin{figure}[b]
\centering
\includegraphics[scale=0.5]{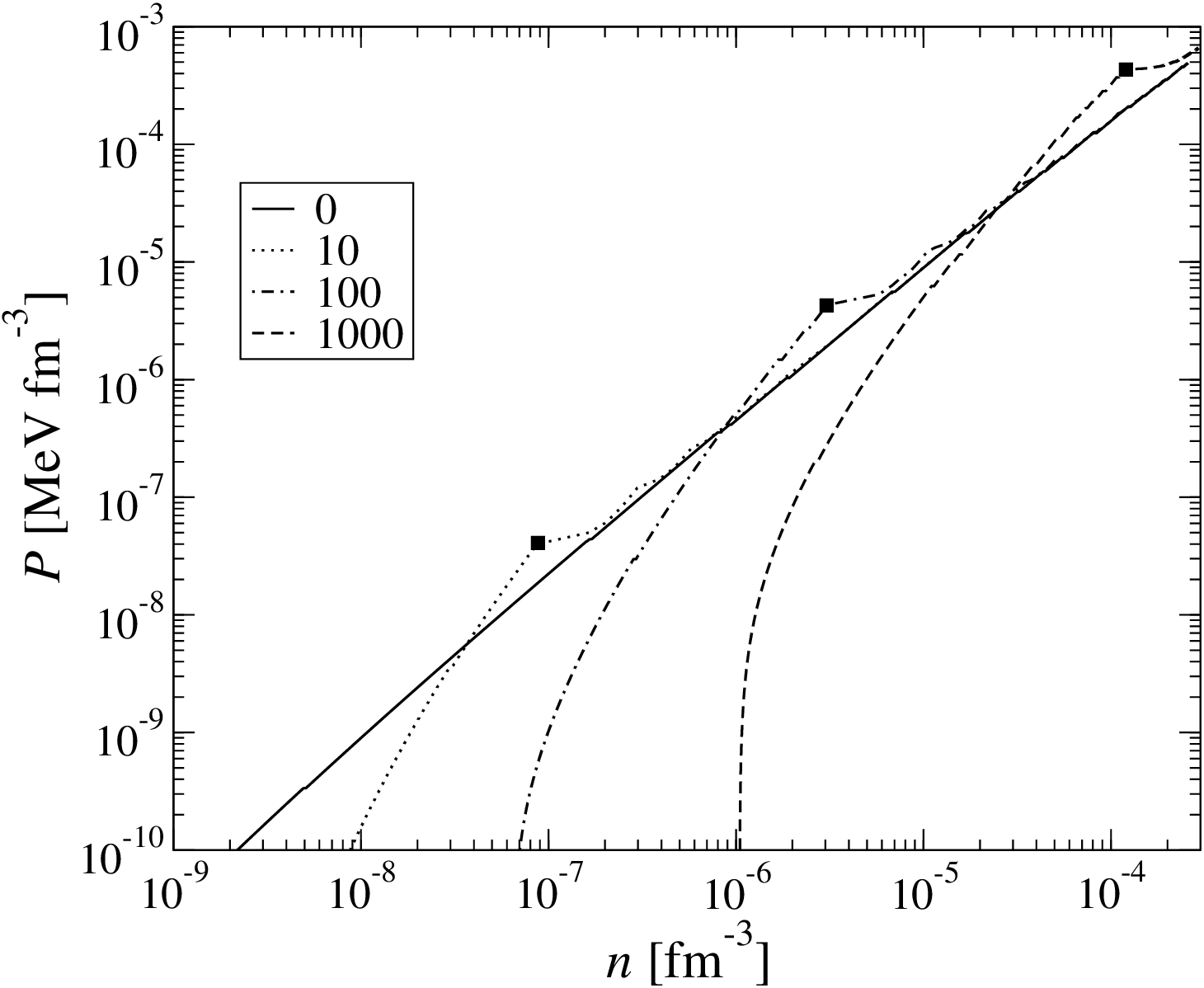}
\caption[]{Pressure $P$ versus average nucleon number density $n$ in the outer crust of a cold non-accreting neutron star for different magnetic 
field strengths $B_\star$. The filled squares indicate the points above which the lowest level $\nu=0$ is fully occupied.}
\label{fig2}
\end{figure}

As shown in Fig.~\ref{fig2}, the strongly quantizing magnetic field prevailing in magnetar interiors is found to have a large impact on the equation 
of state in the regions where only a few Landau levels are filled. In particular, the quantization of electron motion makes the outermost 
layers of the crust almost incompressible, the density remaining essentially unchanged over a wide range of pressures. However, the present model is not
well suited for describing the surface of the star because of the nonuniformity of the electron gas~\cite{lai01}. In addition, at finite temperatures 
thermal effects can considerably change the equation of state~\cite{tho98}. With increasing 
density, the effects of the magnetic field become less and less important as more and more levels are populated and the equation of state 
matches smoothly with that obtained in the absence of magnetic fields. 

\section{Equilibrium composition and equation of state of neutron-star crusts for strongly quantizing magnetic fields}
\label{strong}

A magnetic field is strongly quantizing if only the lowest level $\nu=0$ is filled. This situation occurs whenever the electron 
number density $n_e$ satisfies the inequality (see e.g. Chap. 4 in Ref.\cite{hae07})
\begin{equation}
\label{17}
n_e<\frac{1}{\sqrt{2}\pi^2 a_m^3}
\end{equation}
where $a_m=\sqrt{\hbar c/e B}$. Since the average nucleon density is given by $n=(A/Z)n_e$, Eq.~(\ref{17}) can 
be equivalently expressed as $n < n_B$ with
\begin{equation}
\label{18}
n_B \simeq 1.24\times 10^{-9} \frac{A}{Z} B_\star^{3/2}\, {\rm fm}^{-3}\, .
\end{equation} 

In the following sections, electrons will be assumed to fill only the lowest level $\nu=0$ in all regions of the outer crust. 
We found that this assumption is fulfilled whenever $B_\star>1304$. 

\subsection{Equation of state}
\label{eos}

In strongly quantizing magnetic fields, the electron energy density~(\ref{10}) and the electron pressure~(\ref{11}) 
reduce to 
\begin{equation}
\label{19}
\mathcal{E}_e=\frac{B_\star m_e c^2}{(2 \pi)^2 \lambda_e^3}\psi_+(x_e)-n_e m_e c^2\, ,
\end{equation}
\begin{equation}
\label{20}
P_e=\frac{B_\star m_e c^2}{(2 \pi)^2 \lambda_e^3}\psi_-(x_e)\, ,
\end{equation}
respectively, 
with  
\begin{equation}
\label{21}
x_e=\frac{2\pi^2 \lambda_e^3 n_e}{B_\star}\, .
\end{equation}
The electron chemical potential can be obtained from
\begin{equation}
\label{22}
\gamma_e=\sqrt{1+x_e^2}\, .
\end{equation}

In the upper layers of the crust where $x_e\ll 1$, the electron pressure~(\ref{20}) is approximately given by 
\begin{equation}
\label{23}
P_e\approx \frac{1}{3} m_e c^2 n_e^3 \biggl[\frac{2\pi^2 \lambda_e^3}{B_\star}\biggr]^2\, .
\end{equation}
Substituting Eq.~(\ref{23}) into Eq.(\ref{15}) with $P=0$ using Eq.~(\ref{14}) yields the average density at the surface 
of a cold non-accreting magnetar~\cite{lai91}
\begin{equation}
\label{24}
n_s \approx \frac{A_s}{Z_s}\biggl[\frac{1.44423 Z_s^{2/3} e^2}{m_e c^2}\left(\frac{B_\star}{2\pi^2\lambda_e^3}\right)^2\biggr]^{3/5} \, ,
\end{equation}
with $Z_s$ and $A_s$ the proton number and the charge number of the equilibrium nuclide at the surface.  
Considering that the surface of a neutron star is made of iron with $Z_s=26$ and $A_s=56$ leads to
\begin{equation}
\label{25}
n_s \simeq 2.5\times 10^{-10} B_\star^{6/5}\, {\rm fm}^{-3}\, .
\end{equation}
This simple formula shows that the stronger the magnetic field is, the higher the surface density. It should be stressed 
however that Eq.~(\ref{25}) provides only an approximate estimate of the surface density because for sufficiently strong fields the 
condition $x_e\ll 1$ is not fulfilled. 
Moreover, the present model is not strictly valid at the surface of a neutron star, as mentioned earlier.  
Using Eq.~(\ref{21}) and (\ref{24}), the condition $x_e\ll 1$ at the neutron star surface translates to
\begin{equation}
\label{26}
B_\star \ll \frac{2 \pi^2}{Z_s^2} \left(\frac{\hbar c}{1.44423 e^2}\right)^3\simeq 2.5\times 10^4\,.
\end{equation}

In the dense region of the outer crust where $n\gg n_s$, the lattice pressure~(\ref{14}) is negligible and the total pressure is 
approximately given by 
\begin{equation}
\label{27}
P\simeq P_e \approx m_e c^2 n_e^2 \frac{\pi^2 \lambda_e^3}{B_\star}\, .
\end{equation}
Inverting this equation yields 
\begin{equation}
\label{28}
n = \frac{A}{Z} \left(\frac{ P B_\star}{m_e c^2 \pi^2 \lambda_e^3}\right)^{1/2} \, .
\end{equation}

\begin{figure}[b]
\centering
\includegraphics[scale=0.5]{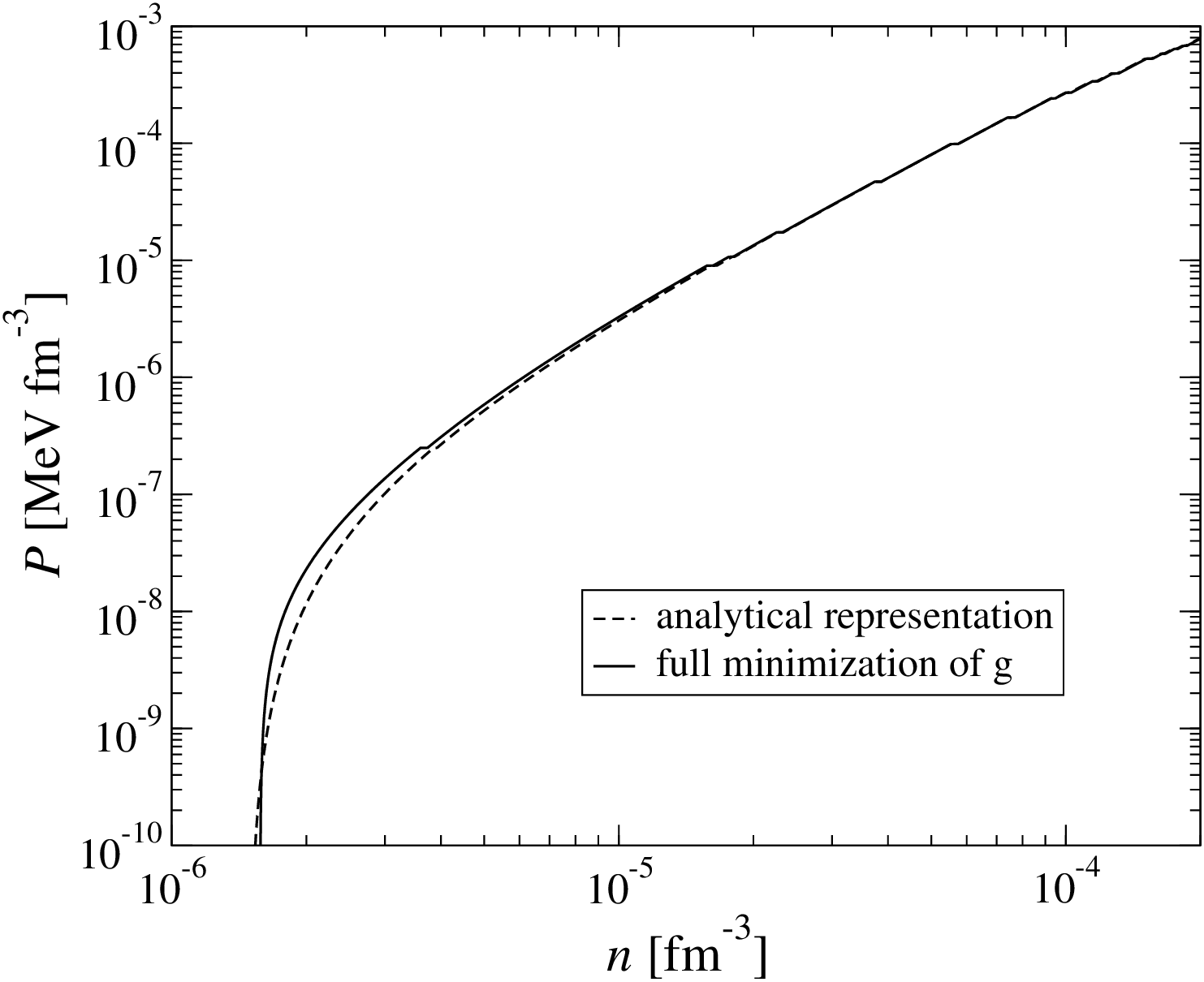}
\caption[]{Pressure $P$ versus average nucleon number density $n$ in the outer crust of a cold non-accreting neutron star for $B_\star=1400$
from the full minimization of the Gibbs free energy (solid line) and from the analytical representation~(\ref{29}) (dashed line).}
\label{fig3}
\end{figure}

Interpolating between the shallow and the deep regions of the outer crust, the density $n$ in a layer at pressure $P$ 
can be approximately expressed as
\begin{equation}
\label{29}
n \approx {n}_s \left(1+\sqrt{\frac{P}{P_0}}\right)\, ,
\end{equation}
where 
\begin{equation}
\label{30}
P_0= m_e c^2 \frac{\bar n_s^2 \pi^2\lambda_e^3}{B_\star} \left(\frac{Z}{A}\right)^2 \simeq 1.82\times 10^{-11} B_\star^{7/5}\left(\frac{Z}{A}\right)^2 \, {\rm MeV\, fm}^{-3}.
\end{equation}
As illustrated in Fig.~\ref{fig3}, the analytical representation~(\ref{29}) yields a fairly good fit to the equation of state 
obtained from the full minimization of the Gibbs free energy $g$. The typical error  is found to be less than $11 \%$ for $B_\star>10$ in 
any region of the outer crust where the condition~(\ref{17}) holds.

At densities $n\gg n_B$ or equivalently at pressures $P\gg P_B$, many Landau levels are populated so that the 
quantization effects disappear and the properties of the crust are almost unaffected by the magnetic field. According to 
Eqs.~(\ref{18}) and (\ref{28}), the pressure $P_B$ is given by 
\begin{equation}
\label{PB}
P_B= \frac{B_\star^2}{2 \pi^2} \frac{m_e c^2}{\lambda_e^3} \, . 
\end{equation}

\subsection{Composition}
\label{comp}

The results about the crustal composition presented in Sec.~\ref{results} can be qualitatively understood using a
simplified atomic mass formula. Neglecting Coulomb and surface contributions, the mass of a nucleus with $Z$ protons and $A$ 
nucleons is given by 
\begin{equation}
\label{31}
M^\prime(A,Z)=A(a_v+J(1-2 y_e)^2+m_u c^2) + Z m_e c^2\, ,
\end{equation}
where $y_e\equiv Z/A$ is the electron fraction, $a_v$ is the binding energy of symmetric nuclear matter, $J$ the symmetry energy and 
$m_u$ the atomic mass unit (ignoring here the small difference between neutron and proton masses). For the HFB-21 nuclear mass model 
that we consider here~\cite{gcp10}, $a_v=-16.053$ MeV and $J=30$ MeV. 
Dropping the lattice energy density $\mathcal{E}_L$ (which is a small correction to the total energy density $\mathcal{E}$), the Gibbs free 
energy per nucleon as given by Eq.~(\ref{16}) reduces to
\begin{equation}
\label{32}
g=a_v+J(1-2 y_e)^2+m_u c^2 + y_e \mu_e\, .
\end{equation}
Minimizing Eq.~(\ref{32}) for a given pressure $P\sim P_e$ (i.e. $\mu_e$ fixed) and treating $y_e$ as a continuous variable yields
\begin{equation}
\label{33}
y_e =\frac{1}{2} -\frac{\mu_e}{8 J}\, .
\end{equation}
Using Eqs.~(\ref{21}),(\ref{22}),  and (\ref{27}) leads to
\begin{equation}
\label{34}
y_e = \frac{1}{2}\left(1 -\sqrt{\frac{P}{P_{\rm neu}}}\right)\, ,
\end{equation}
where 
\begin{equation}
\label{35}
P_{\rm neu} =\frac{4 B_\star J^2}{\pi^2 \lambda_e^3 m_e c^2}\, .
\end{equation}

\begin{figure}[b]
\centering
\includegraphics[scale=0.5]{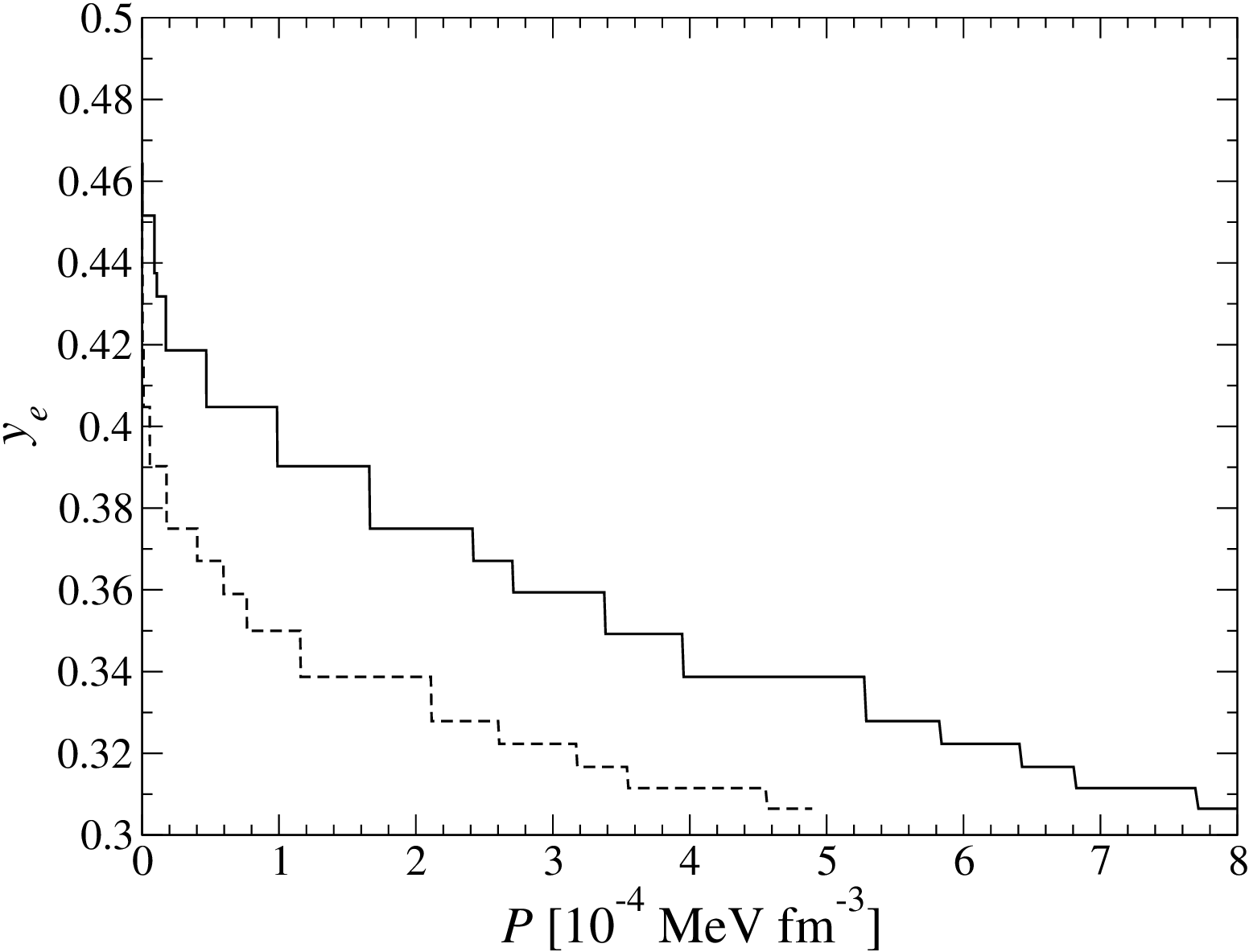}
\caption[]{Electron fraction $y_e$ versus pressure $P$ in the outer crust of a cold non-accreting neutron star for $B_\star=1400$ 
(solid line) and for $B_\star=0$ (dashed line). Note that in both cases the equilibrium nuclide at the bottom of the outer crust 
is $^{124}$Sr even though the neutron-drip pressures are different.}
\label{fig4}
\end{figure}

For comparison, using Eqs.~(\ref{10b}) and (\ref{11b}) the electron fraction in the absence of magnetic field is approximately given 
by 
\begin{equation}
\label{34b}
y_e^0 = \frac{1}{2}\left(1 -\left(\frac{P}{P_{\rm neu}^0}\right)^{1/4}\right)\, ,
\end{equation}
where 
\begin{equation}
\label{35b}
P_{\rm neu}^0 =\left(\frac{4 J}{m_e c^2}\right)^4 \frac{m_e c^2}{12 \pi^2 \lambda_e^3}\, ,
\end{equation}
assuming $P\simeq P_e$ and $x_r\gg 1$. Introducing the isospin asymmetry parameters $\eta=1-2 y_e$ and $\eta^0=1-2 y_e^0$, 
their ratio is given by 
\begin{equation}
\frac{\eta}{\eta^0} =\left(\frac{P}{P_{\rm neu}}\right)^{1/4}\left(\frac{P_{\rm neu}^0}{P_{\rm neu}}\right)^{1/4}\, .
\end{equation}
As will be shown in the next section $P\ll P_{\rm neu}$ in any region of the outer crust. 
Noting that 
\begin{equation}
\left(\frac{P_{\rm neu}^0}{P_{\rm neu}}\right)^{1/4}=\sqrt{\frac{J}{m_e c^2}}\frac{2}{(3 B_\star)^{1/4}}   < 2 
\end{equation}
for $B_\star>1304$ (strongly quantizing field), we find that $\eta < \eta^0$. In other words, nuclei in the outer crust of a 
magnetar are more symmetric than those found in the outer crust of a weakly magnetized neutron star at the \emph{same} pressure. 
This conclusion is confirmed by numerical calculations using the experimental and HFB-21 atomic masses, as shown in Fig.~\ref{fig4}.

\subsection{Neutron-drip transition}
\label{drip}

With increasing pressure, the crustal matter becomes more and more neutron rich, as shown by Eq.~(\ref{34}). The pressure $P_{\rm neu}$
can be interpreted as the pressure at which nuclei will convert to neutron drops. In reality, neutrons start to drip out of nuclei above 
some pressure $P_{\rm drip}$ which is lower than $P_{\rm neu}$. This transition occurs when the neutron chemical potential $\mu_n$ 
exceeds the neutron rest mass energy. As can be seen in Tables~\ref{tab1}-\ref{tab5}, the equilibrium nucleus at the neutron-drip point is 
independent of the magnetic field strength and is found to be $^{124}$Sr for the HFB-21 atomic mass model considered here. 
The reason is the following. Equilibrium with respect to weak interaction processes requires 
\begin{equation}
\label{36}
\mu_p+\mu_e=\mu_n\, ,
\end{equation}
where $\mu_p$ is the proton chemical potential. However at equilibrium, the neutron chemical potential coincides with the Gibbs free 
energy per nucleon $g$. Neglecting the small contribution of the lattice energy density, Eq.~(\ref{16}) leads to
\begin{equation}
\label{37}
\mu_e\approx m_e c^2 + \frac{A}{Z}\left(\mu_n- \frac{M^\prime(A,Z)}{A}\right)\, .
\end{equation}
Substituting the neutron-drip value of the neutron chemical potential $\mu_n=m_n c^2$ in Eqs.(\ref{36}) and (\ref{37}), we find
\begin{equation}
\label{38}
\mu_p-m_p c^2 = Q_{n,\beta}+\frac{A}{Z}\left(\frac{M^\prime(A,Z)}{A}-m_n c^2 \right)\, ,
\end{equation}
where $m_p$ is the proton mass and $Q_{n,\beta}$ is the beta decay energy of the neutron. The quantity on the left-hand side of Eq.~(\ref{38})
is approximately equal to the opposite of the one-proton separation energy. The equilibrium nucleus at neutron drip is therefore 
uniquely determined by atomic masses. Using the two-parameter mass formula discussed in Sec.~\ref{comp}, we find that the 
proton fraction is approximately given by 
\begin{equation}
\label{39}
\frac{Z_{\rm drip}}{A_{\rm drip}}\approx \frac{1}{2}\sqrt{1+\frac{a_v}{J}}\, .
\end{equation}
Substituting the values of $a_v$ and $J$ from the atomic mass model HFB-21 in Eq.~(\ref{39}) yields a fairly good estimate of the proton 
fraction of $^{124}$Sr with an error of about $11\%$ only. Note, however, that without Coulomb and surface terms in the mass 
formula, it is not possible to determine $Z_{\rm drip}$ and $A_{\rm drip}$ separately. 

Equation~(\ref{37}) shows that the electron chemical potential at neutron-drip is independent of the magnetic field and is given by 
\begin{equation}
\label{40}
\mu_e^{\rm drip}=m_e c^2 + \frac{A_{\rm drip}}{Z_{\rm drip}}\left(m_n c^2- \frac{M^\prime(A_{\rm drip},Z_{\rm drip})}{A_{\rm drip}}\right)\, .
\end{equation}
For $^{124}$Sr, we find $\mu_e^{\rm drip}\simeq 26$~MeV. Using Eqs.~(\ref{21}), (\ref{22}), (\ref{27}) and (\ref{40}) implies that the 
pressure at the neutron-drip point increases linearly with the magnetic field strength (in the strongly quantizing regime) as shown in 
Fig.~\ref{fig1} and is given by 
\begin{equation}
\label{41}
P_{\rm drip} = \frac{m_e c^2}{\lambda_e^3} \frac{(\gamma_e^{\rm drip})^2}{4\pi^2} B_\star\, .
\end{equation}
The corresponding baryon density is given by 
\begin{equation}
\label{41b}
n_{\rm drip} = \frac{A_{\rm drip}}{Z_{\rm drip}}\frac{\gamma_e^{\rm drip} }{2\pi^2\lambda_e^3} B_\star\, .
\end{equation}
Using the two-parameter mass formula yields 
\begin{equation}
\label{42}
P_{\rm drip} = P_{\rm neu}\left(1-\sqrt{1+\frac{a_v}{J}}\right)^2 < P_{\rm neu}\, ,
\end{equation}
\begin{equation}
\label{42b}
{n}_{\rm drip}=\frac{4 J B_\star}{\pi^2 \lambda_e^3 m_ec^2}\sqrt{1+\frac{a_v}{J}}^{-1}\left(1-\sqrt{1+\frac{a_v}{J}}\right)\, .
\end{equation}

Electrons fill only the lowest level $\nu=0$ in any region of the outer crust provided $P_B\geq P_{\rm drip}$. 
Using Eqs.~(\ref{PB}) and (\ref{41}), we find that this condition is equivalent to $B_\star>B_\star^{\rm drip}$ 
with 
\begin{equation}
\label{43}
B_\star^{\rm drip}=\frac{1}{2}(\gamma_e^{\rm drip})^2 \, .
\end{equation}
This estimate could have been immediately obtained from Eq.~(\ref{8}) requiring $x_e^2\geq 0$. 
For $^{124}$Sr, we find $B_\star^{\rm drip}\simeq 1300$. 

In the absence of magnetic fields, the neutron-drip pressure and baryon  density (in the ultrarelativistic regime $x_r \gg 1$) 
are approximately given by 
\begin{equation}\label{41c}
P_{\rm drip}^0 \approx \frac{m_e c^2}{\lambda_e^3} \frac{(\gamma_e^{\rm drip})^4}{12 \pi^2}
\end{equation}
\begin{equation}\label{41d}
n_{\rm drip}^0 \approx \frac{A_{\rm drip}}{Z_{\rm drip}}\frac{(\gamma_e^{\rm drip})^3 }{3\pi^2\lambda_e^3} \, ,
\end{equation}
respectively. Using Eqs.~(\ref{41}), (\ref{43}) and (\ref{41c}) leads to 
\begin{equation}\label{44}
\frac{P_{\rm drip}}{P_{\rm drip}^0} = \frac{3}{2} \frac{B_\star}{B_\star^{\rm drip}}>\frac{3}{2}\, .
\end{equation}
This shows that the neutron-drip transition occurs at a higher pressure in a magnetar than in a weakly magnetized 
neutron star.  

\subsection{Elastic properties}
\label{elastic}

Because a sufficiently strong magnetic field changes the composition of the outer crust of a neutron star, it can also have an 
impact on the crustal properties. In view of the recent detection of QPOs in the x-ray flux of giant flares from SGRs, a 
particularly important property of strongly magnetized neutron star crusts is the shear modulus which determines the frequencies 
of torsional oscillations. 

\begin{figure}[b]
\centering
\includegraphics[scale=0.5]{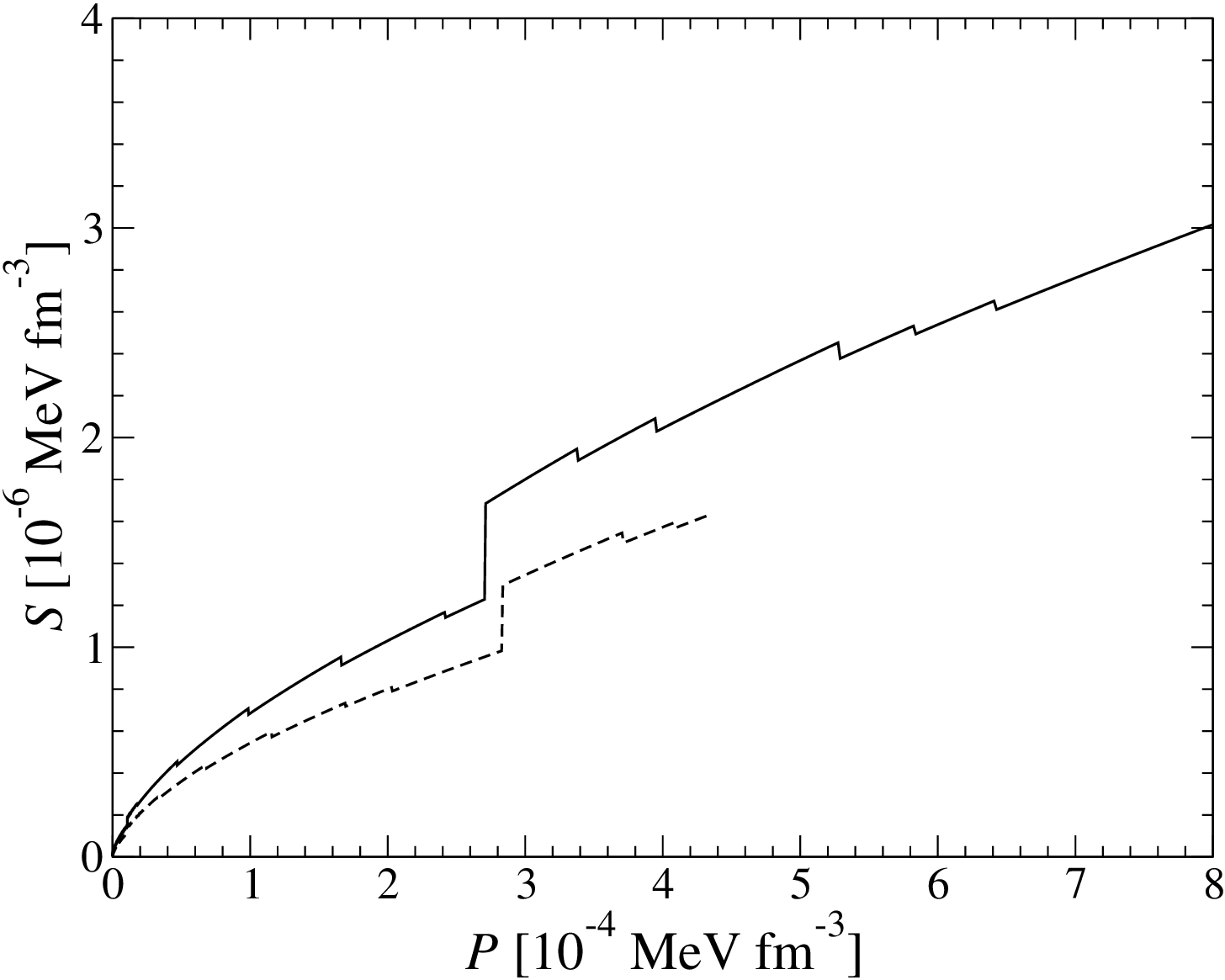}
\caption[]{Effective shear modulus $S$ versus pressure $P$ in the outer crust of a cold non-accreting neutron star for $B_\star=1400$ 
(solid line) and for $B_\star=0$ (dashed line). Note that the neutron-drip pressures are different in the two cases.}
\label{fig5}
\end{figure}

We have calculated the ``effective'' shear modulus $S$ of the outer crust, assuming that it is made of a body-centered-cubic 
lattice polycrystal, using the following expression~\cite{ogi90}~: 
\begin{equation}
S = 0.1194 n_{\rm N} \frac{Z^2 e^2}{R_{\rm N}} \, ,
\end{equation} 
where $R_{\rm N}$ is the ion-sphere radius defined by 
\begin{equation}
R_{\rm N}=\left(\frac{3}{4\pi n_{\rm N}}\right)^{1/3} \, .
\end{equation} 
As shown in Fig.~\ref{fig5}, the effective shear modulus of the outer crust of a neutron star can be enhanced by the presence of 
a strong magnetic field. 

\subsection{Global structure}
\label{struc}

In the outer crust of a non-rotating neutron star of gravitational mass $\mathcal{M}$ and circumferential radius $R$, 
the general relativistic equations can be approximately written as (see, e.g., Ref.~\cite{pea11})
\begin{equation}\label{45}
\frac{dP}{dz} \approx g_s \rho   \,  ,
\end{equation}
where $g_s$ is the surface gravity defined by
\begin{equation}\label{46}
g_s = \frac{G\mathcal{M}}{R^2}\left(1 - \frac{r_g}{R}\right)^{-1/2}
\end{equation}
$z$ is the depth below the surface, $r_g=2 G\mathcal{M}/c^2$ is the Schwarzschild radius and $\rho\approx n m_u$ is the 
mass density. 
The gravitational mass $\Delta \mathcal{M}\ll \mathcal{M}$ contained in the outer crust is approximately given by
\begin{equation}\label{47}
\Delta \mathcal{M} \approx \frac{8\pi R^3 P_{\rm drip}}{c^2}\left(\frac{R}{r_g}-1\right)\, .
\end{equation}
Comparing Eqs.~(\ref{41}) and (\ref{41c}) shows that the crustal mass for neutron stars endowed with strongly quantizing 
magnetic fields is larger than that of weakly magnetized neutron stars with the \emph{same} mass and radius, and is given by 
\begin{equation}\label{48}
\Delta \mathcal{M} =\frac{2 R^3 m_e (\gamma_e^{\rm drip})^2 B_\star}{\pi \lambda_e^3}\left(\frac{R}{r_g}-1\right)\, .
\end{equation}
Likewise the magnetic field increases the baryonic mass contained in the outer crust, which is approximately given by 
\begin{equation}\label{49}
\Delta M_B \approx \sqrt{1-\frac{r_g}{R}}\Delta \mathcal{M}\, .
\end{equation}
On the contrary, the depth $z$ below the surface where neutron drip occurs and which therefore delimits the boundary between 
the outer and inner crusts, does not depend on the magnetic field strength. 
Indeed, in the absence of magnetic fields assuming that the main contribution to the pressure is due to ultra-relativistic 
electrons (i.e., $P\propto \rho^{4/3}$), Eq.~(\ref{45}) can be easily solved, leading to~\cite{pea11} 
\begin{equation}\label{50}
z^0\approx\frac{8 P^0_{\rm drip} R}{\rho^0_{\rm drip} c^ 2}\sqrt{\frac{R}{r_g}\left(\frac{R}{r_g}-1\right)}\, ,
\end{equation}
which can be expressed as 
\begin{equation}\label{51}
z^0=\frac{2 m_e \gamma_e^{\rm drip} y_e^{\rm drip} R }{m_u}\sqrt{\frac{R}{r_g}\left(\frac{R}{r_g}-1\right)}\, ,
\end{equation}
where we used Eqs.~(\ref{41c}) and (\ref{41d}). In the presence of a strongly quantizing magnetic field,  
Eq.~(\ref{27}) shows that the pressure varies approximately as $P\propto \rho^2$. Solving Eq.(\ref{45}) thus yields
\begin{equation}\label{52}
z\approx\frac{4 P_{\rm drip} R}{\rho_{\rm drip} c^ 2}\sqrt{\frac{R}{r_g}\left(\frac{R}{r_g}-1\right)}\, .
\end{equation}
Using Eqs.~(\ref{41}) and (\ref{41b}) leads to Eq.~(\ref{51}) so that $z=z^0$. 

\section{Conclusion}
We calculated the composition and the equation of state of the outer crust of cold non-accreting neutron 
stars endowed with very strong magnetic fields of order $B\gg m_e^2 c^3/(e\hbar)\simeq 4.4\times 10^{13}\, \rm G$, as 
measured in soft-gamma ray repeaters, anomalous x-ray pulsars and even in a few radio pulsars~\cite{mcgill,ka11}. For this purpose, 
we made use of the most recent experimental atomic mass data~\cite{audi11} complemented with the latest Hartree-Fock-Bogoliubov 
atomic mass model~\cite{gcp10}. 

The Landau quantization of electron motion due to the strong magnetic field is found to have a significant impact on the 
neutron-star crust properties: 
(i) it changes the crustal composition (the sequence of 
equilibrium nuclides being different than that found in weakly magnetized crusts, as summarized in Table~\ref{tab6}) and (ii) it 
makes the matter less neutron rich as shown in Fig.~\ref{fig4} and tends to prevent neutrons from dripping out of nuclei (the pressure 
at neutron drip increasing with $B$ as shown in Fig.~\ref{fig1}). As a consequence, the presence of a strong magnetic field can have an 
impact on the crustal properties like the shear modulus, as shown in Fig.~\ref{fig5}. 
These results may have implications for the interpretation of the quasiperiodic oscillations observed in soft gamma-ray repeaters. 
Likewise, other crustal properties such as the 
thermal and electric conductivities could be affected. The present results might therefore also impact the thermal and magnetic field 
evolution of magnetars. This warrants further study. 

The outer crust of a magnetar is also found to be much more massive than the outer crust of a weakly magnetized neutron star with the 
\emph{same} gravitational mass $\mathcal{M}$ and circumferential radius $R$. This implies that the contribution of magnetars to the 
galactic enrichment in nuclides heavier than iron from the rapid neutron capture process ($r$ process) of nucleosynthesis following the 
ejection and the decompression of crustal material~\cite{gcjp11}, could be much more important than previously thought.

\section*{Acknowledgments}
The present work was supported by the bilateral project between FNRS (Belgium), Wallonie-Bruxelles-International (Belgium) and the 
Bulgarian Academy of Sciences. This work was also supported by NSERC (Canada) and CompStar, a Research Networking Programme of the 
European Science Foundation.

\end{document}